\documentclass[aip,superscriptaddress,preprint,apl]{revtex4}
\usepackage{amsmath,amsfonts,amssymb,natbib}
\usepackage[english]{babel}
\usepackage{booktabs,longtable}
\usepackage{graphicx}
\usepackage{dcolumn}
\usepackage{bm}
\usepackage[T1]{fontenc}
\usepackage[latin9]{inputenc}
\usepackage{color}
\usepackage{textcomp}
\usepackage{amsmath}

\begin{document}

\title{Magnetotransport in Bi$_2$Se$_3$ thin films epitaxially grown on Ge(111)}

\author{T. Guillet}
\affiliation{Univ. Grenoble Alpes, CEA, CNRS, Grenoble INP (Institute of Engineering Univ. Grenoble Alpes), INAC-Spintec, 38000 Grenoble, France}

\author{A. Marty}
\affiliation{Univ. Grenoble Alpes, CEA, CNRS, Grenoble INP (Institute of Engineering Univ. Grenoble Alpes), INAC-Spintec, 38000 Grenoble, France}

\author{C. Beign\'e}
\affiliation{Univ. Grenoble Alpes, CEA, CNRS, Grenoble INP (Institute of Engineering Univ. Grenoble Alpes), INAC-Spintec, 38000 Grenoble, France}

\author{C. Vergnaud}
\affiliation{Univ. Grenoble Alpes, CEA, CNRS, Grenoble INP (Institute of Engineering Univ. Grenoble Alpes), INAC-Spintec, 38000 Grenoble, France}

\author{M.-T. Dau}
\affiliation{Univ. Grenoble Alpes, CEA, CNRS, Grenoble INP (Institute of Engineering Univ. Grenoble Alpes), INAC-Spintec, 38000 Grenoble, France}

\author{P. No$\ddot{\rm e}$l}
\affiliation{Univ. Grenoble Alpes, CEA, CNRS, Grenoble INP (Institute of Engineering Univ. Grenoble Alpes), INAC-Spintec, 38000 Grenoble, France}

\author{J. Frigerio}
\affiliation{LNESS-Dipartimento di Fisica, Politecnico di Milano, Polo di Como, via Anzani 42, 22100 Como, Italy }

\author{G. Isella}
\affiliation{LNESS-Dipartimento di Fisica, Politecnico di Milano, Polo di Como, via Anzani 42, 22100 Como, Italy }

\author{M. Jamet}
\affiliation{Univ. Grenoble Alpes, CEA, CNRS, Grenoble INP (Institute of Engineering Univ. Grenoble Alpes), INAC-Spintec, 38000 Grenoble, France}

\date{\today}

\begin{abstract}
Topological insulators (TIs) like Bi$_2$Se$_3$ are a class of material with topologically protected surface states in which spin-momentum locking may enable spin-polarized and defect-tolerant transport. In this work, we achieved the epitaxial growth of Bi$_2$Se$_3$ thin films on germanium, which is a key material for microelectronics. Germanium also exhibits interesting properties with respect to the electron spin such as a spin diffusion length of several micrometers at room temperature. By growing Bi$_2$Se$_3$ on germanium, we aim at combining the long spin diffusion length of Ge with the spin-momentum locking at the surface of Bi$_2$Se$_3$. We first performed a thorough structural analysis of Bi$_2$Se$_3$ films using electron and x-ray diffraction as well as atomic force microscopy. Then, magnetotransport measurements at low temperature showed the signature of weak antilocalization as a result of two-dimensional transport in the topologically protected surface states of Bi$_2$Se$_3$. Interestingly, the magnetotransport measurements also point out that the conduction channel can be tuned between the Bi$_2$Se$_3$ film and the Ge layer underneath by means of the bias voltage or the applied magnetic field. This result suggests that the Bi$_2$Se$_3$/Ge junction is a promising candidate for tuning spin-related phenomena at interfaces between TIs and semiconductors.   
\end{abstract}


\maketitle

\section{Introduction}

In the past decade, topological insulators (TI) have gained much interest in the field of spintronics for the generation and detection of spin currents. Three-dimensional (3D) TI are predicted to exhibit original properties like topologically protected surface states (TSS) showing Dirac band dispersion and strong spin-momentum locking \cite{Hasan2010,Zhang2009}. The existence of these states was rapidly confirmed by angle-resolved photoemission spectroscopy (ARPES) \cite{Hsieh2009} on films grown by molecular beam epitaxy (MBE). Since then, the growth and transport properties of bismuth-based compounds such as Bi$_2$Se$_3$ and  Bi$_2$Te$_3$ \cite{Wu2011,Kun2011,Lee2018} were extensively studied both experimentally    
 \cite{Banerjee2016,Wang2014,Wang2016,Chatterjee2015,Wang2013} and theoretically \cite{Shen2011,Shen2011b,Shen2010}. 
In particular, the existence of topologically protected surface states could be demonstrated through their signature in magnetotransport experiments. Following the predictions of two-dimensional transport, quantum corrections to the conductivity are expected and low temperature magnetoresistance measurements exhibit weak (anti)localization (WL and WAL) 
 \cite{HLN1980}. 
Furthermore, the surface states exhibit a helical spin texture due to strong spin-momentum locking. Hence, a charge current flowing into the surface states is spin-polarized. The epitaxial growth of TI on conventional semiconductors appears as a very promising route to develop original spintronic devices by coupling spin generation in TSS at the interface with a long spin diffusion length ($l_{sf}$) material that can also be optically active like germanium \cite{Ciccacci2016,Jamet2017}. In this study, we report the epitaxial growth of a Bi$_2$Se$_3$ thin film by MBE on low-doped Ge(111) ($p\approx$ $10^{15}$ cm$^{-3}$). Magnetoresistance measurements at low temperature clearly show the two-dimensional transport in the TSS of Bi$_2$Se$_3$. Since the film is $n$-doped by growth-induced selenium vacancies, we demonstrate the existence of a $pn$ junction that can exhibit two transport regimes depending on the bias voltage and magnetic field applied to the junction. By adjusting both parameters, it is possible to select the Bi$_2$Se$_3$ channel with spin-momentum locking at surface states or the Ge channel with long spin diffusion length.


\section{Sample growth}

\begin{figure}[h!]
\begin{center}
\includegraphics[width=\textwidth]{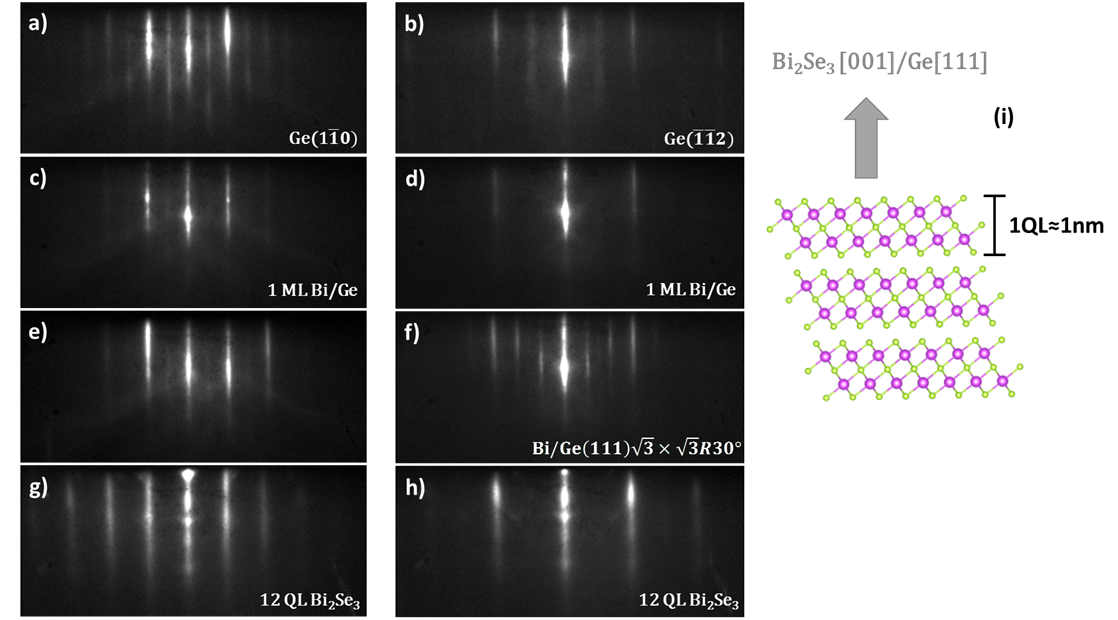} \\
\end{center}
\caption{RHEED patterns recorded during the growth of Bi$_2$Se$_3$ on Ge(111). (a), (b) Bare Ge(2$\times$8) reconstructed surface after ion bombardment and annealing up to 850$^{\circ}$C.  (c), (d) After deposition of one monolayer of Bi at room temperature. (e), (f) Bi/Ge(111)-($\sqrt{3}\times\sqrt{3}$)R30$^{\circ}$  surface reconstruction after annealing at 500$^{\circ}$C during 10 minutes. (g), (h) 12 QL of Bi$_2$Se$_3$ grown at 220$^{\circ}$C. (i) Quintuple layer structure of Bi$_2$Se$_3$.} \label{fig1}
\end{figure}

Ultrathin films of  Bi$_2$Se$_3$ were grown on Ge(111) by MBE, the surface quality and structure were followed by reflection high-energy electron diffraction (RHEED) all along the growth. Before the growth of Bi$_2$Se$_3$, the Ge(111) surface was first annealed up to 850$^{\circ}$C under ultrahigh vacuum (UHV) ($p\approx$5$\times$10$^{-10}$ mbar) in order the remove the native germanium oxide. Then, we used soft argon etching and performed a subsequent annealing to obtain the Ge (2$\times$8) surface reconstruction as shown in Fig.~\ref{fig1}a and ~\ref{fig1}b. In order to initiate the epitaxial growth of Bi$_2$Se$_3$, we first deposited  one monolayer (ML) of Bi at room temperature (see Fig.~\ref{fig1}c and ~\ref{fig1}d) and annealed the substrate until the  Bi/Ge(111)-($\sqrt{3}\times\sqrt{3}$)R30$^{\circ}$  surface reconstruction appeared as shown in Fig.~\ref{fig1}e and ~\ref{fig1}f. This Bi layer prevents the reaction of Ge with Se to form GeSe alloys \cite{Lee2018}. Bi$_2$Se$_3$ was then grown by co-depositing Bi and Se at a substrate temperature of 220$^{\circ}$C. Bi and Se were evaporated using an e-beam evaporator and a Knudsen cell operating at $\approx$200$^{\circ}$C, respectively. Bi and Se evaporation rates were adjusted in order to reach a high Se:Bi ratio of about 15:1 and limit the presence of Se vacancies in the film. Fig.~\ref{fig1}g and ~\ref{fig1}h show characteristic RHEED patterns along two different azimuths of the as-grown 12 quintuple layers (QL) of Bi$_2$Se$_3$. 1 QL corresponds to 1 nm. The lamellar crystal structure is schematically shown in Fig.~\ref{fig1}i. 

\begin{figure}[h!]
\begin{center}
\includegraphics[width=\textwidth]{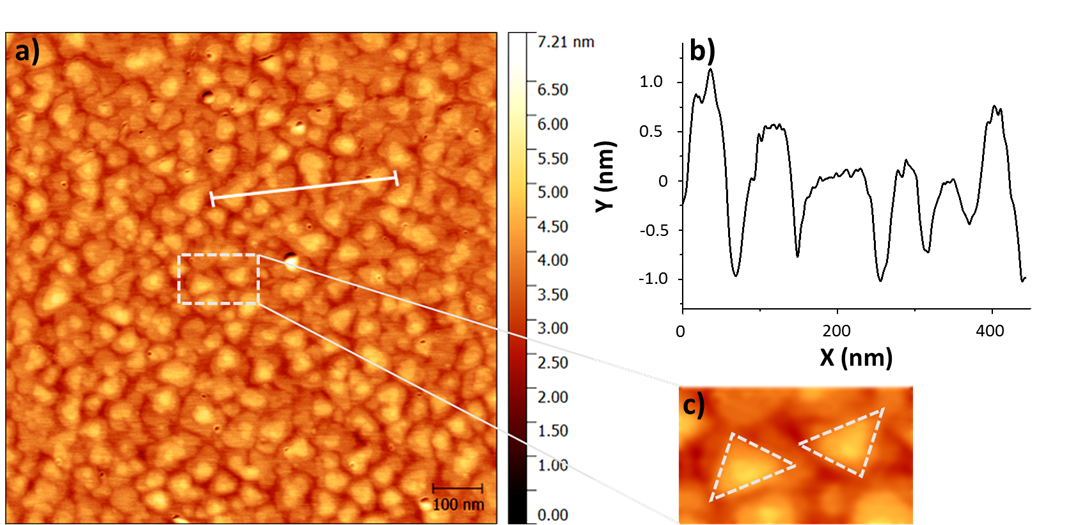} \\
\end{center}
\caption{(a) Atomic force microscopy image of the 12 QL-thick Bi$_2$Se$_3$ film. (b) Height profile along the light grey solid line shown in (a). (c) Zoom-in of the AFM image showing two triangular grains pointing at opposite directions.} \label{fig2}
\end{figure}

A characteristic atomic force microscopy (AFM) image in Fig.~\ref{fig2}a shows the film morphology. Typical step high of 1QL can be seen on the height profile shown in  Fig.~\ref{fig2}b. The root-mean-square roughness is of the order of 0.51 nm. The film is capped with 2 nm of aluminum to prevent from oxidation. The aluminum layer is grown in two steps: 1 nm was deposited by e-beam evaporation  and 1 nm by magnetron sputtering in the same UHV setup. The final RHEED pattern exhibits rings characteristic of a polycrystalline Al layer. Since the total thickness is 12 QL, we expect Bi$_2$Se$_3$ to exhibit gapless TSS \cite{Hsieh2009}. \\


\section{Structural characterization}


In-plane and out-of-plane x-ray diffraction (XRD) measurements were performed with two different diffractometers. The grazing incidence
X-ray diffraction (GIXD) was done with a SmartLab Rigaku diffractometer equipped with a copper rotating anode beam tube ($K_{\ensuremath{\alpha}}$=1.54 \AA) operating at 45 kV and 200 mA. Parallel in-plane collimators of 0.5$^{\circ}$ of resolution were used both on the source and detector sides. The
out-of-plane diffraction was performed using a Panalytical Empyrean diffractometer equipped with a cobalt source, ($K_{\ensuremath{\alpha}}$=1.79 \AA)
operated at 35 kV and 50 mA. The incident beam divergence slit was set at 0.125$^{\circ}$ and the diffracted beam was measured using a camera PIXcel-3D detector allowing a resolution of 0.02$^{\circ}$ per pixel. Both diffractometers are equipped with multilayer mirrors on the incident beam and $K_{\beta}$ filter on the diffracted beam.

\begin{figure}[h!]
\begin{center}
\includegraphics[width=\textwidth]{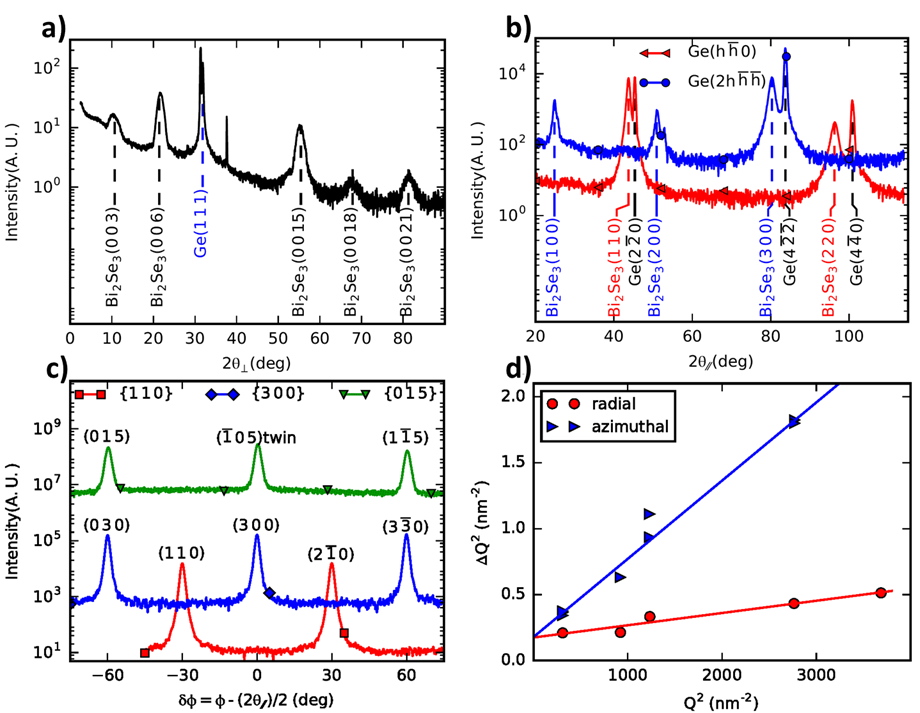} \\
\end{center}
\caption{(a) Out-of-plane symmetric $\theta/2\theta$ spectrum close to the substrate
reciprocal direction Ge$(h\,h\,h)$ substrate. A small offset $\delta\omega=\omega-\theta=0.25^{\circ}$ was used to attenuate the substrate peaks and reduce the contribution from parasitic peaks ($K_{\beta}$, $\lambda/2$...) still present despite the $K_{\beta}$ filtering. (b) In-plane GIXD radial scans along the two reciprocal directions separated from each other by 30$^{\circ}$: Ge$(h\,\bar{h}\,0)$ and Ge$(2h\,\bar{h}\,\bar{h})$. The incidence angle was optimized to $0.32^{\circ}$. (c) In-plane GIXD azimuthal scans for two peaks families \{300\} and \{110\}. The spectra of \{100\}, \{200\} and \{220\} families were also measured but not shown here for clarity. The azimuthal scan of the \{015\} reflection shows the presence of twin domains. This measurement was performed using the same grazing incidence, 0.32$^{\circ}$, but with suitable exit angles. (d) Square of the FWHM of Bragg peaks (in $Q$ units) as a function of the square of the momentum transfer $Q$ for both in-plane radial and azimuthal scans. The two linear fits show equal intercepts at the origin giving an estimation of the domains size: $D\approx$15 nm.} \label{fig3}
\end{figure}

Fig.~\ref{fig3}a shows the symmetric out-of-plane $\theta/2\theta$ diffraction spectra along the Ge$(h\,h\,h)$ reciprocal direction. In addition
to the substrate Ge(111) peak, the 5 other peaks can be attributed to the rhombohedral structure $R\bar{3}m$ of Bi$_2$Se$_3$ \cite{nakajima1963crystal}. They are indexed in the hexagonal unit cell ( a=0.4143 nm and c=2.8636 nm) which consists of three Se-Bi-Se-Bi-Se quintuple layers separated from each other by a van der Waals gap. The relative intensities of the peaks are consistent with the calculated structure factors. It can be noticed that the 2 most intense peaks in this reciprocal direction can be easily understood : the (0015) reflection corresponds to the average lattice spacing c/15 of the 15 atomic planes regardless the nature of the atoms, whereas the (006) reflection is attributed to the average lattice spacing separating the 6 Bi atomic layers, Bi having the larger diffusion factor. 

Figures ~\ref{fig3}b and ~\ref{fig3}c show GIXD measured with an optimized incidence angle of $0.32\text{$^{\circ}$}$. Radial scans (Fig.~\ref{fig3}b) along the two 30$^{\circ}$ apart in-plane directions Ge$(2h\,\bar{h}\,\bar{h)}$ and Ge$(h\,\bar{h}\,0$) give the epitaxial relationship between Bi$_2$Se$_3$ and the Ge substrate : Bi$_2$Se$_3$(110)$\Vert$Ge(1$\bar{1}$0). The peak positions corresponding to the ones of bulk Bi$_2$Se$_3$ show that there is no substrate induced in-plane strain. Azimuthal scans around the Bi$_2$Se$_3$(hk0) Bragg peaks (Fig.~\ref{fig3}c) show the in-plane alignment of Bi$_2$Se$_3$ and Ge crystals: 30$^{\circ}$ rotational domains are completely absent. However, pure in-plane measurements cannot exclude twinning which generally occurs due to the simultaneous nucleation of twinned domains on lattice mismatched substrates \cite{kriegner2017twin}. Indeed, the ABCAB and ACBAC stackings of the quintuple layer structure give in-plane diffraction peaks \{hk0\} at the same positions. Nevertheless, the 3-fold symmetry of the out-of-plane \{015\} reflexions allows to quantify the degree of twinning \cite{bonell2017growth}. The measurement shows that the film is composed of both twins in equal proportions which leads to the presence of triangular grains pointing at opposite directions as shown in Fig.~\ref{fig2}c.

From the Bragg peaks width in radial and azimuthal scans as a function of the momentum transfer: $Q=\frac{4\pi}{\lambda}sin(\theta)$ , we can estimate the in-plane domain size $D$, the in-plane mosaicity $\Delta\xi$ and the lattice parameter distribution, $\Delta a/a$ according to the quadratic relations \cite{renaud1999growth}:
\begin{equation}
\Delta Q_{rad}^{2}=\left(\frac{2\pi}{D}\right)^{2}+Q^{2}\left(\frac{\Delta a}{a}\right)^{2}
\end{equation}
\begin{equation}
\Delta Q_{azi}^{2}=\left(\frac{2\pi}{D}\right)^{2}+Q^{2}\Delta\xi^{2}
\end{equation}

where the radial and azimuthal full width at half maximum (FWHM) in $Q$ units are related to the diffraction peaks widths through the relations: $\Delta Q_{rad}=\frac{4\pi}{\lambda}cos(\theta)\frac{\Delta\left(2\theta\right)}{2}$ and $\Delta Q_{azi}=Q\Delta\Phi$. Both least-squares fits in Fig.~\ref{fig3}d give similar domain sizes close to $D$=15 nm which is a lower bound since we did not consider here the setup resolution. The slopes of the fits give a lattice parameter dispersion less than 1 \% (radial) and an in-plane mosaicity of $\Delta\xi$=1.4$^{\circ}$ (azimuthal) which are rather low values considering the presence of twinned domains \cite{bonell2017growth}. Despite their weak intensity, the presence of forbidden peaks like (100) and (200) of Bi$_2$Se$_3$ in the radial scan of Fig.~\ref{fig3}b comes from the non-integer number of unit cells in crystal grains. This can be due to a non-uniform layer thickness (12$\pm$1 QL), surface roughness and a substantial twin boundaries density. The absence of thickness fringes around Bragg peaks in the out-of-plane measurements (Fig.~\ref{fig3}a) can also be explained by the film roughness shown in the AFM image of Fig.~\ref{fig2}a. 


\section{Magnetotransport}

\begin{figure}[h!]
\begin{center}
\includegraphics[width=\textwidth]{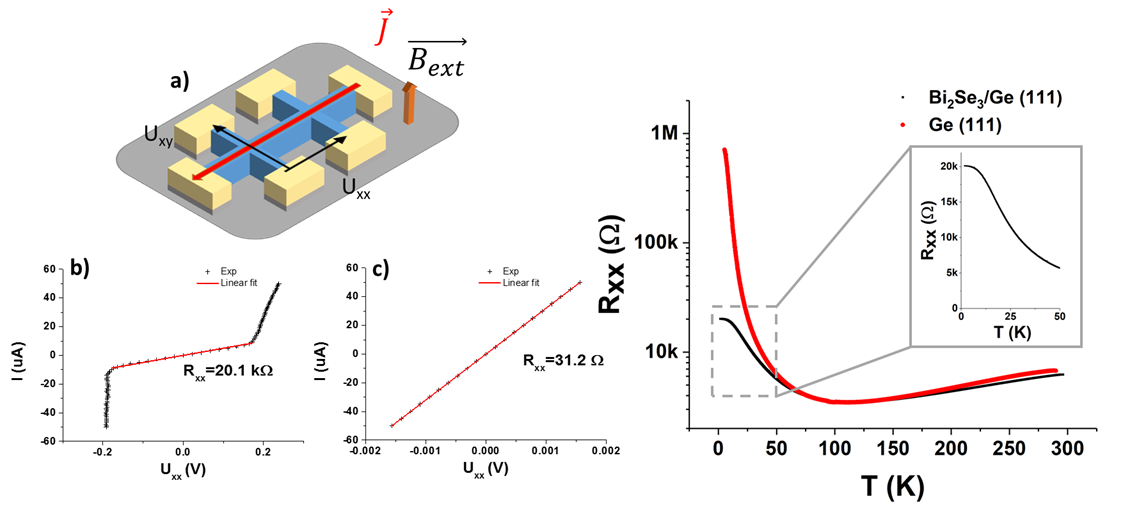} \\
\end{center}
\caption{(a) Sketch of the Hall bar microstructure and description of the 4-probe measurements geometry. (b), (c) $I(V)$ curves recorded between the two extreme current leads at 1.6 K and 295 K respectively. (d) Log-scale representation of the longitudinal resistance $R_{xx}$ as a function of temperature, the red curve is for Bi$_2$Se$_3$/Ge (111) and the black curve for the substrate. Inset: zoom-in at low temperature showing the resistance saturation.} \label{fig4}
\end{figure}
\begin{figure}[h!]
\begin{center}
\includegraphics[width=\textwidth]{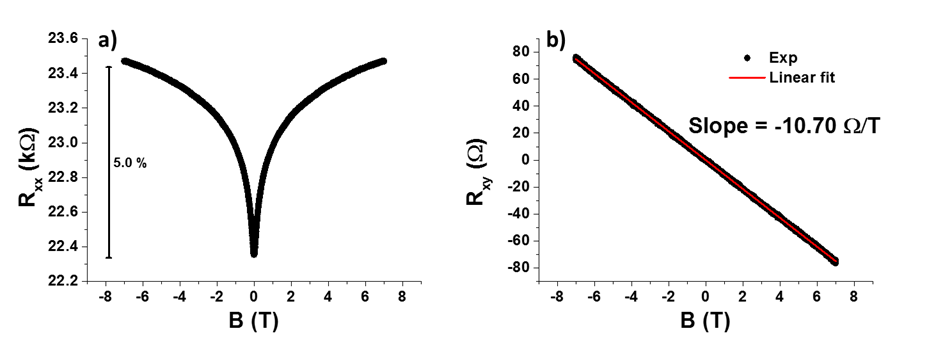} \\
\end{center}
\caption{(a) $R_{xx}$ as a function of the applied magnetic field at 1.6 K. The field is perpendicular to the film. (b) Corresponding transverse Hall resistance  $R_{xy}$ at 1.6K.} \label{fig5}
\end{figure}

As-grown Bi$_2$Se$_3$ films were patterned into micron-sized  Hall bars (dimensions : length $L$=130 $\mu$m and width $W$=10 $\mu$m)  as shown in Fig.~\ref{fig4}a to perform magnetoresistance (MR) and Hall measurements. The microfabrication of Hall bars required three successive steps : laser lithography to define the pattern, e-beam evaporation of Au(90 nm)/Ti(5 nm) ohmic contacts and ion beam etching.
 MR and Hall measurements were performed in a helium closed cycle cryostat working in the 1.6-295 K temperature range and equipped with a superconducting magnet delivering up to 7 Tesla. 
 Fig.~\ref{fig4}b and ~\ref{fig4}c show current-voltage ($I(V)$) curves recorded between the two extreme current leads at 1.6 K and 295 K respectively. The maximum applied current is 50 $\mu$A. We clearly see the rectification effect at the Bi$_2$Se$_3$/Ge interface at low temperature. Part of the current is shunted into Ge for $U_{xx}\geqslant$ 0.2 V. For $U_{xx}<$ 0.2 V, the linear behavior indicates the ohmic character of the Ti/Au contacts on Bi$_2$Se$_3$. The red curve in Fig.~\ref{fig4}d shows the temperature dependence of the DC 4-probe longitudinal resistance $R_{xx}$=$U_{xx}/I$ for an applied current of 10 $\mu$A.
We find an overall semiconducting character due to the current shunting into the Ge substrate. However, the resistance saturation at low temperature shown in the inset of Fig.~\ref{fig4}d corresponds to electrical transport into the Bi$_2$Se$_3$ film due to the increasingly high resistance of the Ge substrate and Bi$_2$Se$_3$/Ge contacts below 10 K as shown in  Fig.~\ref{fig4}b . 
As a comparison, we show in Fig.~\ref{fig4}d (black curve) the temperature dependence of the Ge substrate resistance. As shown in Fig.~\ref{fig4}a and in the following , the current shunting into the Ge substrate can be restored at low temperatures by applying larger current \textit{i.e.} larger bias voltages. Fig.~\ref{fig5}a and ~\ref{fig5}b show MR and Hall measurements recorded at 1.6 K with an applied current of 1 $\mu$A.
$R_{xx}$ clearly exibits a MR dip at low magnetic field corresponding to WAL. 
The linear dependence of $R_{xy}$ on the magnetic field is interpreted in terms of a single-carrier electrical transport. From the slope, we obtain a $n$-type doping as expected for MBE-grown Bi$_2$Se$_3$ films where Se vacancies act as donors. We find a carrier concentration of 4.6$\times$10$^{19}$cm$^{-3}$ assuming 3D transport (both into surface states and the bulk) and 5.4$\times$10$^{13}$cm$^{-2}$ if we consider 2D transport into the surface states. We further find a low mobility of 37 cm$^2$/(V.s) which might be explained by the high concentration of twin defects as unveiled by XRD and AFM measurements. \\

\begin{figure}[h!]
\begin{center}
\includegraphics[width=\textwidth]{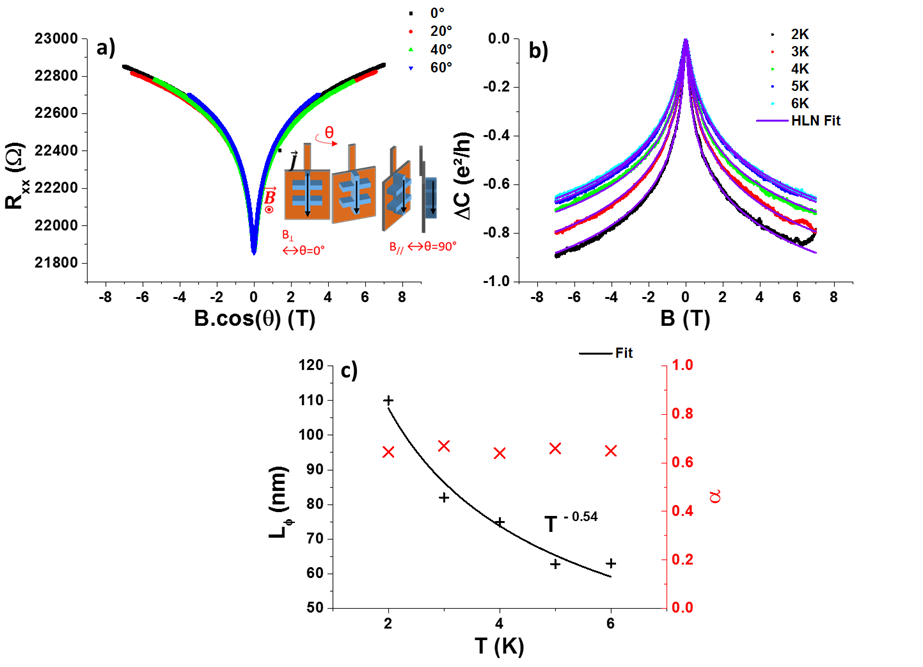} \\
\end{center}
\caption{(a) $R_{xx}$ as a function of $Bcos(\theta)$. The angle $\theta$ is defined in the inset. (b) Magnetoconductance $\Delta C$ normalized to the quantum of conductance $e^2/h$ as a function of temperature. Red solid lines are fits using the HLN model. (c) Parameters extracted from the HLN model: $L_\phi$ is the effective phase coherence length and $\alpha$, the characteristic parameter related to the number of transport channels.} \label{fig6}
\end{figure}

The observation of WAL strongly suggests a 2D electrical transport into TSS \cite{Shen2011,HLN1980}. This is supported by the temperature and angular dependences of the magnetoresistance. Fig.~\ref{fig6}a presents MR measurements as a function of the projected magnetic field $Bcos(\theta)$. All the curves perfectly overlap at low fields which is the signature of WAL and a sign of robust topological transport \cite{Chen2014}. Fig.~\ref{fig6}b  shows the film magnetoconductance at temperatures varying from 2 K to 6 K. In Fig.~\ref{fig6}c, the data are fitted using the Hikami-Larkin-Nagaoka (HLN) two-dimensionnal quantum diffusion model \cite{HLN1980,Wang2014,Chatterjee2015}:
 
\begin{equation}
\Delta C = -\frac{\alpha e^2}{2\pi^2\hbar}\left[ \psi\left(\frac{\hbar}{4eBL_{\phi}^2}+\frac{1}{2}\right)-ln\left(\frac{\hbar}{4eBL_{\phi}^2}\right)   \right]
\end{equation} 

where $\psi$ is the digamma function, $B$ is the applied magnetic field perpendicular to the film, $L_{\Phi}$ is the effective phase coherence length and $\alpha$ a parameter related to the number of channels contributing to the transport. $\alpha=0.5$ is for one surface contributing to the transport and $\alpha=1$ for two surfaces contributing. In the literature, $\alpha$ varies from 0.25 to 1 depending on the thickness \cite{Oh2011}, or the film fabrication technique \cite{Wang2016}. Using the HLN model, we can extract a temperature independent $\alpha$ value of 0.61. In this intermediate case where $0.5<\alpha<1$, we estimate that the 2D transport mostly takes place at the bottom surface in contact with Ge, with a smaller contribution from the top surface due to its roughness (see Fig.~\ref{fig2}a)  . Interestingly, our measurements show no sign of weak localisation even at 1.6 K. Advanced theoretical models demonstrated that WL is expected in ultrathin films where bulk states are quantized along the film normal. Hence, a pure WAL signature (without WL) supports the fact that the electrical transport takes place into TSS and not into 2D quantized bulk states \cite{Shen2010,Shen2010b,Shen2011b,Wang2013}. $L_{\phi}$ decreases with increasing the temperature as $T^{-0.54}$, which is in good agreement with the theory predicting $L_{\phi}\propto T^{-0.50}$. \\

\begin{figure}[h!]
\begin{center}
\includegraphics[width=\textwidth]{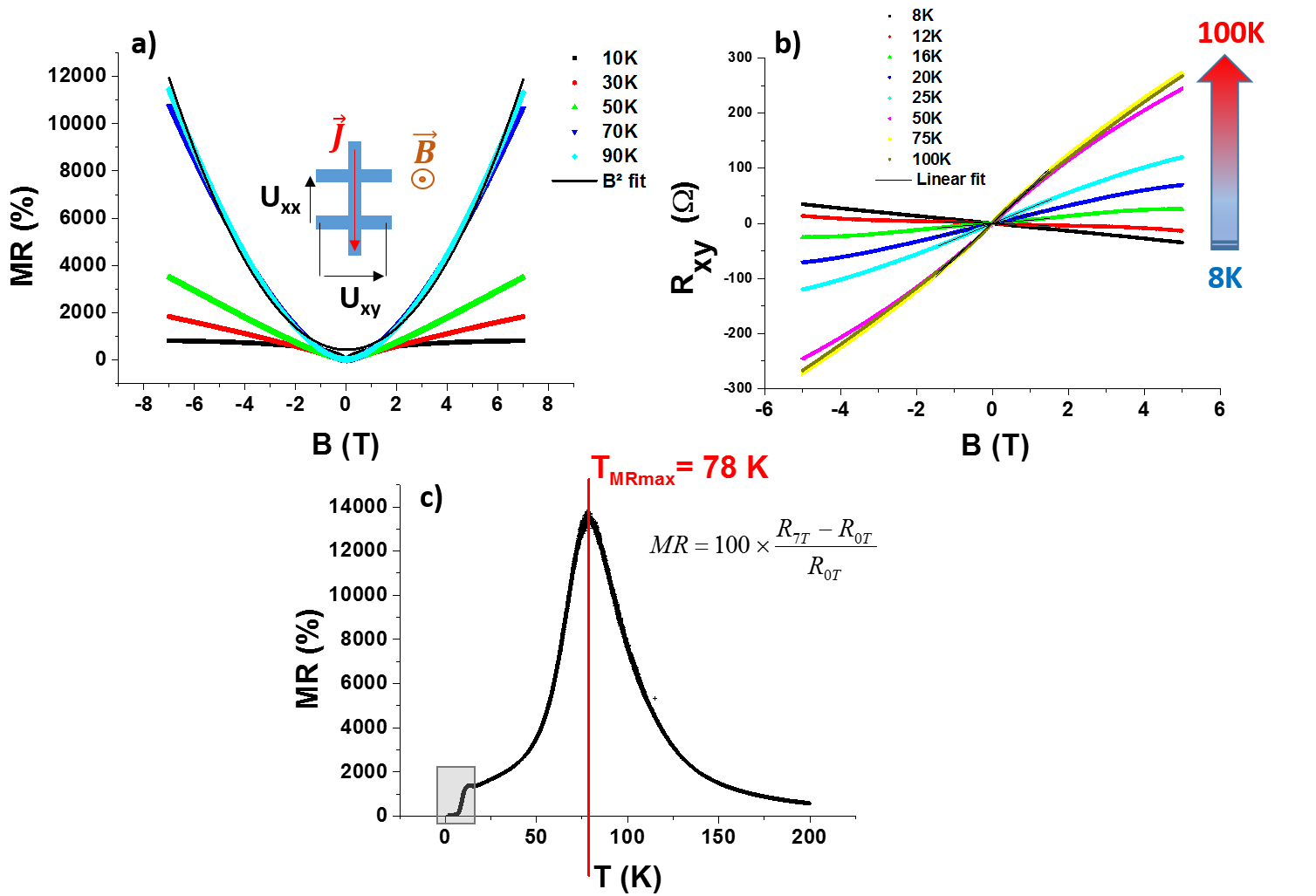} \\
\end{center}
\caption{(a) Longitudinal magnetoresistance measured at different temperatures. The inset shows the measurement geometry ($B$ being perpendicular to the film). (b) Transerve resistance $R_{xy}$ in the same geometry as a function of the temperature, the sign of the Hall effect changes as the Ge channel is being activated  (c) Continuous MR(T) curve. The grey box shows the transition to magnetotransport in the Bi$_2$Se$_3$ film at low temperature.} \label{fig7}
\end{figure}

Fig.~\ref{fig7}a shows MR measurements at higher temperatures, where the germanium conducting channel is thermally activated. In this case, we find a conventional Lorentz MR behavior where $\frac{\Delta R}{R_0} \propto (\mu B)^2$ characteristic of a 3D bulk transport. The magnitude of this MR is the one expected from the high carrier mobility in germanium (see http://www.ioffe.ru/SVA/NSM/Semicond/Ge/electric.html). Fig.~\ref{fig7}b shows $R_{xy}$ as a function of temperature in Hall configuration. When the temperature increases, the sign of the Hall effect changes from negative for $n$-type doping (Bi$_2$Se$_3$ carriers) to positive for $p$-type doping (Ge carriers). By measuring continuously the longitudinal resistance $R_{xx}$ as a function of the temperature for 0 Tesla and 7 Tesla, we could extract a continuous MR curve given by: $\frac{ R_{7T}- R_{0T}}{ R_{0T}}$ and shown in Fig.~\ref{fig7}c. It shows a maximum at 78 K corresponding to the temperature at which all the dopants in Ge are thermally activated and the electron-phonon scattering is minimum. For $T\leq$10 K, we observe a sharp drop of the MR when the charge current is no more shunted into the Ge substrate but only flows into the Bi$_2$Se$_3$ film where the MR is limited to some percents (see Fig.~\ref{fig5}a). \\ 

\begin{figure}[h!]
\begin{center}
\includegraphics[width=\textwidth]{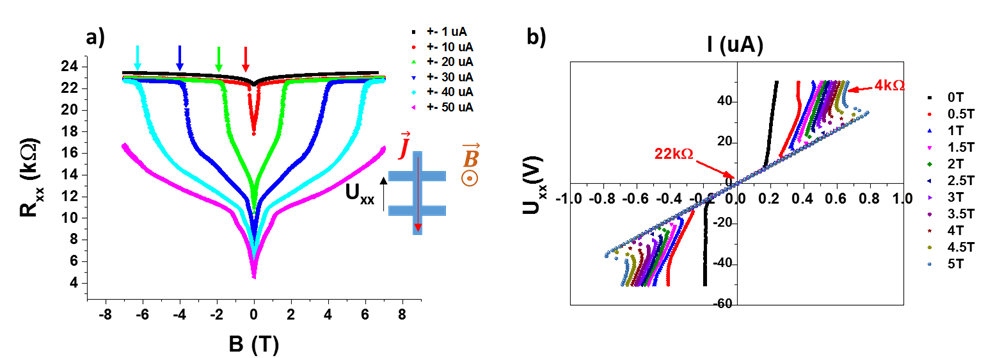} \\
\end{center}
\caption{(a) MR curves measured at 1.6 K for different applied currents. Two magnetotransport regimes can be clearly identified. The transition magnetic field is indicated by arrows. The inset shows the measurement geometry ($B$ being perpendicular to the film). (b) $I(V)$ curves at different magnetic fields exhibiting a $pn$-junction behavior. } \label{fig8}
\end{figure}

Fig.~\ref{fig8}a shows MR measurements at 1.6 K for bias currents varying from 1 $\mu$A to 50 $\mu$A. To eliminate the offset voltage due to thermal effects, the current sign is changed from $+I$ to $-I$ and the longitudinal resistance is calculated using: $R_{xx}=\frac{R_{+I}+ R_{-I}}{2}$. Two different transport regimes can be observed depending on the applied magnetic field. The critical field separating those two regimes (indicated by arrows in Fig.~\ref{fig8}a) increases with the bias current. By measuring $I(V)$ curves at different magnetic fields in Fig.~\ref{fig8}b, we find the characteristic magnetic field dependence of a $pn$-junction $I(V)$ curve \cite{Holloway1975,Wan2011} in parallel with a resistor. The $n$-doped (resp. $p$-doped) layer can be associated to the Bi$_2$Se$_3$ film (resp. the germanium substrate). When the current is kept low enough ($\approx$8 $\mu$A), the $I(V)$ curve keeps a ohmic character and the magnetotransport takes place in the Bi$_2$Se$_3$ film only. For a current higher than 8 $\mu$A, the current source generates a high enough bias voltage to make the $pn$-junction conducting and the current mostly flows into germanium. Despite its high resistivity (~1 k$\Omega$cm at 2 K), the germanium substrate is so thick (350 $\mu$m) that its resistance is much lower than the one of the Bi$_2$Se$_3$ film. This regime corresponds to the steeper slope in the $I(V)$ curve. The diode threshold voltage $V_d$ increases with the applied magnetic field. Hence, for a given bias voltage, the magnetic field allows to select the conducting channel and magnetotransport properties. In regime 1 where the applied magnetic field is lower than a critical field (marked by a vertical arrow in Fig.~\ref{fig8}a), the current flows in the germanium substrate and the resistance (resp. MR) is low (resp. high). In this case, the MR curve is the one of the Ge substrate that we measured independently using a second device made of pure Ge with Ti/Au ohmic contacts (not shown). In regime 2 where the applied magnetic field is higher than the critical field, the current flows in the Bi$_2$Se$_3$ film and the resistance (resp. MR) is high (resp. low). At the transition between the two regimes, we observe very sharp steps in MR curves in Fig.~\ref{fig8}a with slopes up to 20 $\Omega$/mT and negative differential resistances in $I(V)$ curves (Fig.~\ref{fig8}b). These phenomena require further investigation and are out of the scope of this work. \\
 

This $pn$-junction effect at a semiconductor/topological insulator could be of great interest to tune spin transport since one can control whether the charge current is spin-polarized (regime 2) or not (regime 1). It also paves the way to develop spin-FET structures where the spin information can be transmitted by the application of an electric field. 
Finally, we stress the fact that we obtained the same magnetotransport results using a 2 $\mu$m-thick epitaxial germanium film (of comparable $p$-type doping) on semi-insulating Si(111) instead of a germanium substrate. This Ge-on-Si epilayer was deposited on a 3-inch high-resistivity Si(111) wafer by low-energy plasma enhanced chemical vapor deposition at a deposition rate of $\approx$4 nm/s and a substrate temperature of 500$^{\circ}$C \cite{Gatti2014}. Post-growth annealing cycles have been used to reduce the threading dislocation density down to $\approx$2$\times$10$^7$ cm$^{-2}$ and to improve the crystal quality \cite{Osmond2009}. Due to the much higher resistance of the 2 $\mu$m-thick Ge channel as compared to the one of the Ge wafer (by almost two orders of magnitude), the transition between regime 1 and regime 2 occurs at higher bias voltages and magnetic fields.


\section{Conclusion}
In this work, we have successfully grown by epitaxy a 12 QL-thick Bi$_2$Se$_3$ film on germanium. Despite the presence of twin boundaries, we obtained a high-crystalline quality material with a low surface roughness of $\pm$1 QL. Low temperature magnetotransport measurements showed the signature of two-dimensional weak antilocalization in the topological surface states of Bi$_2$Se$_3$ with a phase coherence length of the order of 110 nm at 2 K. By studying the temperature dependence of the MR and Hall effect, we found that the electrical current flows into the Bi$_2$Se$_3$ film at low temperature. In this case, we measured a low MR and $n$-type doping. When the temperature increases, the electrical current is progressively shunted into the Ge layer and we measured a high MR and $p$-type doping. Finally, at 1.6 K, we could tune the conduction channel between Bi$_2$Se$_3$ and Ge by adjusting the bias voltage or the applied magnetic field. Hence, it could be possible to select electrically or magnetically the Bi$_2$Se$_3$ conduction channel with spin-momentum locking or the Ge conduction channel with a long spin diffusion length. These findings pave the way to design innovative spintronic devices by combining semiconductors and topological insulators for which the energy barrier between the two materials acts as a controllable switch between two spin transport regimes.


\section{Acknowledgements}

This work was supported by the French Agence Nationale de la Recherche through the project ANR-16-CE24-0017 TOP-RISE. The LANEF framework (ANR-10-LABX-51-01) is acknowledged for its support with mutualized infrastructure. Partial support is acknowledged to the Horizon-2020 FET  microSPIRE project, ID: 766955.


\newpage

\end{document}